\documentstyle[12pt]{article}
\newcommand{\bc}{\begin{center}}
\newcommand{\ec}{\end{center}}
\newcommand{\be}{\begin{eqnarray}}
\newcommand{\ee}{\end{eqnarray}}
\newcommand{\br}{\begin{right}}
\newcommand{\er}{\end{right}}

\begin{document}
\pagestyle{empty}
\input epsf

\vbox{\vglue 4cm}
\bc
{\Large \bf LEP
       {\boldmath $e^{+}\,e^{-}\,\rightarrow\,\mu^{+}\,
\mu^{-}\,\gamma\,\gamma$}
       events and  their \\
       consequences at future {\boldmath $e^+e^-$} colliders.}
\ec

\vspace{4cm}
\bc{\bf V.A.~Litvin}   \\
{\it Moscow Institute of Physics and Technology   \\
     Dolgoprudny, Moscow region 141700, RUSSIA},
\vspace{0.5cm}

and

\vspace{0.5cm}
{\bf S.R.Slabospitsky}  \\
{\it Institute for High Energy Physics  \\
 Protvino, Moscow Region 142284, RUSSIA}
\ec
\vspace{2cm}
\bc
{\bf Abstract}
\ec
    The $e^{+}\,e^{-}\,\rightarrow\,l^{+}\, \l^{-}\,\gamma\,\gamma$ anomalous
 events, detected by $L3$ de\-tec\-tor at $e^+ e^-$ $CERN-LEP$ collider
have been analysed. It has been shown that the interpretation of such events
as a manifestation of scalar (pseudoscalar) resonance with the mass of 60 GeV
contradicts other experimental data. In case of a possible existence
that kind of resonance, the perspectives to discover one in some processes
at future $e^+e^-$ colliders have been discussed.

\newpage
\section{\bf INTRODUCTION}

  Four unusual events in the reaction
\begin{eqnarray}
e^{+}\,e^{-}\,\rightarrow \,\l^{+}\ \l^{-}\,\gamma\,\gamma \nonumber
\end{eqnarray}
have recently been detected in the experiment $L3$ at the $e^+e^-$
$LEP$ collider at $CERN$ \cite{1}.
The two photon invariant mass in all these events is about 60 GeV.
After original paper \cite{1} the search for analogous events
of process (1) was performed with other detectors at the LEP collider
$[1-8]$.

	As has already been mentioned in \cite{11,12}, the interpretation
of these events meets with some difficulties.

In this paper we analyze the events of process (1),
assuming the existence of a scalar (pseudoscalar)
 resonance R with the mass approximately
equal to 60 GeV and at more general
assumptions about its interactions with photons and Z--bosons,
than \cite{11}. As in previous paper \cite{11} we do
not discuss the nature of such resonance.

	Besides main process (1) other processes, which gave us some
constraints on the coupling constants, have been analyzed. After this
joint analysis of various experiments, the maximum value of the contribution
of this resonance in the cross section of process (1) has been obtained.

	 The perspectives of discovering this hypothetic
resonance in various processes at future $e^+e^-$--colliders also have been
considered.

The paper is organized as follows. The resonance interactions with
 photons and $Z$--bosons and analysis of various experiments at LEP
are considered in Section 2.
The possibilities to discover this resonance at $e^+e^-$--colliders
 are considered in
Section $3$.
The main obtained results are given in the Conclusion.

\section{\bf Extended model.}
\subsection{ Theoretical estimates}

 As has already been mentioned in the Introduction we assume the
existence of a scalar (pseudoscalar)  resonance $R$ with the mass of about 60
GeV.

Proceeding from the requirement of the Lorentz and gauge invariance, the
general form of the
$R\gamma\gamma$,$R\gamma Z$,$RZZ$ interaction will be as follows \cite{11}:
\begin{eqnarray}
R^+ \gamma\gamma & = &  \frac{ g_{\gamma\gamma} }{m_{R}}
 (g^{\mu\nu}(k_{1}k_{2})-k_{1}^{\nu}k_{2}^{\mu})
 e_{1}^{\mu}e_{2}^{\nu},    \nonumber\\
R^{-}\gamma\gamma & = & \frac{g_{\gamma\gamma}}{m_{R}}
 \varepsilon^{\mu\nu\alpha\beta}k_{1}^{\mu}k_{2}^{\nu}
 e_{1}^{\alpha}e_{2}^{\beta},  \nonumber\\
R^{\pm} \gamma Z & = &  \frac{ g_{\gamma z1} }{m_{R}}
 (g^{\mu\nu}(k_{1}k_{2})-k_{1}^{\nu}k_{2}^{\mu})
 e^{\mu}V^{\nu}+\frac{g_{\gamma z2}}{m_R}
 \varepsilon^{\mu\nu\alpha\beta}k_{1}^{\mu}k_{2}^{\nu}
 e^{\alpha}V^{\beta},  \nonumber\\
R^{\pm} ZZ & = &  g_{zz1} m_{Z}  g^{\mu\nu} V_1^{\mu}V_2^{\nu}+
\frac{g_{zz2}}{m_R}k_1^{\nu}k_2^{\mu}V_1^{\mu}V_2^{\nu}+
\frac{g_{zz3}}{m_R} \varepsilon^{\mu\nu\alpha\beta}k_{1}^{\mu}k_{2}^{\nu}
 V_{1}^{\alpha}V_{2}^{\beta}, \nonumber
\end{eqnarray}
  where $R^+(R^-)$ denotes scalar (pseudoscalar) resonance;
$k_1$ and $k_2$ are momenta of two photons (photon and $Z$--boson);
$e^{\nu}\,(V^{\nu})$ is the photon ($Z$--boson) polarization vector.
 The factor $1/m_R$ is introduced for the dimensionless coupling constant.
%
%

Proceeding from the interaction vertices of the R resonance with
photon and $Z$--boson, mentioned above, one can easily
calculate the cross section for various
processes of the R--production.
To determine the value of each effective coupling
constant we will consider different processes which may
result in the appearance of this R--resonance. Moreover, we
will consider the processes where the R--resonance decays into
$\gamma\gamma$ final state. Thus the cross sections of these
processes will depend on all constants, as well as on
$Br(R\,\to\,\gamma\gamma)$ -- the probability of the R decay into
$\gamma\gamma$ :
\begin{eqnarray}
\sigma = f(g_{\gamma\gamma}\,;\,g_{\gamma z1}\,;\,g_{\gamma z2}\,;\,g_{zz1}
\,;\,g_{zz2}\,;\,g_{zz3})Br(R\,\to\,\gamma\gamma) \nonumber
\end{eqnarray}
where $Br(\,R\,\rightarrow\,\gamma\,\gamma) \,=
\, \Gamma(\,R\,\rightarrow\,\gamma\,\gamma\,) \, / \, \Gamma_{tot}(R)$.

	In fact we consider the following reactions:
\be
e^+e^-\,\to\,\mu^+\mu^- R(\,\to\,\gamma\gamma) 	\\
e^+e^-\,\to\,\nu \bar \nu R(\,\to\,\gamma\gamma) 	\\
e^+e^-\,\to\,q \bar q R(\,\to\,\gamma\gamma) 	\\
e^+e^-\,\to\,e^+ e^- R(\,\to\,\gamma\gamma) 	\\
Z\,\to\,\gamma R\,\to\,\gamma\gamma\gamma
\ee

\bc
{\it Processes $e^+e^-\,\to\,f \bar f\gamma\gamma$.}
\ec

The cross sections of reactions (1)  --  (3) (with
$\gamma^{\ast}$ or $Z^{\ast}$ in the s--channel)
were calculated in \cite{11}. An analogous final state
$f\bar f\gamma\gamma$ can be produced as a result
of the following processes:
\be
e^+e^-\,\to\,\gamma R(\,\to\,\gamma Z^{\ast}(\,\to\,f \bar f))
\ee
However, the contribution of this type of diagrams
is negligibly small because of the smallness of the probability for
the decay $ R\,\to\,\gamma Z^{\ast}(\,\to\,f \bar f)$ as compared
with the decay $R\,\to\,\gamma\gamma$.
\begin{center}
{\it Two photon annihilation process}
\end{center}
\begin{eqnarray}
	e^{+}\,e^{-}\,\rightarrow\,e^{+}\,e^{-}\,R
         \,(\,\rightarrow\,\gamma\,\gamma\,) \nonumber
\end{eqnarray}

 A considerable contribution
to this process will be made by the diagrams with photon exchange
in $t$--channel, and $Z$--boson contribution to the $s$ channel
will have some effects, but for the estimates we may
consider two photon approximation.

   The weak interaction contribution from  the remaining diagrams
is negligibly small, as
compared with the QED contribution. Then in the equivalent photon
approximation we have \cite{9}:

\[
 \sigma(e^+ e^- \to e^+ e^- R) \, = \,
        {\eta}^2\frac{8{\pi}^2 \,\Gamma(R \to \gamma \gamma)}
         {s \, m_R} \,f(\frac{m_R^2}{s}),
\]
where $\eta \, = \frac{\alpha}{2\pi}\ln(\frac{s}{4m_e^2}); \quad
 f(\omega) \, = \, \frac{1}{\omega}\,((2+\omega)^2\,\ln(\frac{1}
        {\omega})-2(1-\omega)(3+\omega)).$

\noindent A corresponding expression for the cross section of process (4) is
the following:
\begin{eqnarray}
 \sigma(e^{+}\,e^{-}\,\rightarrow\,e^{+}\,e^{-}\,R\,
	(\rightarrow\,\gamma\,\gamma))\,=\,\sigma(e^{+}\,
        e^{-}\,\rightarrow\,e^{+}\,e^{-}\,R) \,Br(R\,\rightarrow\,
	\gamma\,\gamma).
\end{eqnarray}
The decay width of the $R$--resonance into two photons has the form:

\begin{eqnarray}
 \Gamma(R^{\pm}\,\rightarrow\,\gamma\,\gamma)\,=\,
 \frac{g_{\gamma\gamma}^{2}}{64\pi}m_{R}.
\end{eqnarray}

\begin{center}
{\it $Z$--boson decay into $3\gamma$}
\end{center}

Following constraint of the coupling constants may be derived from the
experimental
values of the branching for the $Z \,\rightarrow \,3\gamma$ decay \cite{9}:
\begin{eqnarray}
\Gamma(Z\,\rightarrow\,\gamma\,R(\,\rightarrow\,\gamma\,\gamma))
\,=\,\Gamma(Z\,\rightarrow\,\gamma\,R)\,Br(R\,\rightarrow\,
\gamma\,\gamma)\,\leq\,\Gamma(Z\,\rightarrow\,3\gamma).
\end{eqnarray}
where the expression for $\Gamma(Z\,\rightarrow\,\gamma\,R)$ is as follows~:
\begin{eqnarray}
\Gamma(Z\,\rightarrow\,\gamma\,R^+)\,=\,
\Gamma(Z\,\rightarrow\,\gamma\,R^-)\,=\,
\frac{(g_{\gamma z1}^2\,+\,g_{\gamma z2}^2)\,
m_Z\,(1\,-\,m_R^2/m_Z^2)^3}{96\pi(m_R^2/m_Z^2)},
\end{eqnarray}

\subsection{ Experimental data analysis}

	As has already been mentioned in the Introduction,
after original paper \cite{1} a search for analogous
events of process (1) was performed with other detectors
at the LEP collider
 $[1-8]$. The following selection criteria
have been used in this search:
\begin{enumerate}
\item [1)] for $\mu^+\mu^-\gamma\gamma$:
\begin{itemize}
\item  $E_{\gamma}\,>\,5$ GeV,
$20^\circ\,<\,\theta\,<\,160^\circ$,
\item  $\varphi_{\gamma l}\,>\,5^\circ$,
 $(m_{\mu^+\mu^-})_{min}\,=\,18$ GeV.
\end{itemize}
\item [2)] for $\nu \bar \nu\gamma\gamma$:
\begin{itemize}
\item  $E_{\gamma}\,<\,5$ GeV,
 $20^\circ\,<\,\theta\,<\,160^\circ$.
\end{itemize}
\item [3)] for $ q \bar q \gamma\gamma$:
\begin{itemize}
\item $E_{\gamma}\,>\,5$ GeV,
 $20^\circ\,<\,\theta\,<\,160^\circ$,
\item $\varphi_{\gamma q}\,>\,25^\circ$,
 $(m_{q\bar q})_{min}\,=\,1$ GeV for $q\,=\,u,d,s$,
\item $(m_{q\bar q})_{min}\,=\,4$ GeV for $q\,=\,c$,
 $(m_{q\bar q})_{min}\,=\,10$ GeV for $ q\,=\,b$.
\end{itemize}
\end{enumerate}
where $\varphi_{\gamma l}$ ($\varphi_{\gamma q}$) is the
minimum isolation angle
between any photon and any lepton (quark); $E_{\gamma}$ is the
minimum energy of any photon; $\theta$ is the polar angle between
photons and charged particles; $(m_{ff})_{min}$ is the  minimum of the
invariant mass pair of final fermions.

	All experimental data $[1-8]$ have been summarized
in Table 1:
%
%

\begin{center}
{\bf Table 1.} Summarized experimental data $[1-8]$.
\end{center}
\vspace{0.5cm}
\begin{center}
\begin{tabular}{|c||c|c|c|}
\hline
\multicolumn{1}{|c||}{\bf Collaboration } & \multicolumn{3}{c|}
{     $\mu\mu\gamma\gamma$     }                   \\ \hline
 & \multicolumn{1}{c|}{$\int {\cal L} d\tau$ , {\it pb}$^{-1}$}
 & \multicolumn{1}{c|}{ N, events} &
\multicolumn{1}{c|}{$M_{\gamma\gamma}$, GeV}    \\ \hline
ALEPH & 43 & 1 & 59.4$\pm$0.2                 \\  \hline
DELPHI& 40 & 1 & 59.0$\pm$2.5                 \\    \hline
L3    & 40 & 3 & 58.8$\pm$0.6                 \\   \cline{4-4}
      &    &   & 59.0$\pm$0.6                 \\   \cline{4-4}
      &    &   & 62.0$\pm$0.6                 \\   \hline
OPAL  & 43 & 0 & ---                          \\    \hline
\end{tabular}
\end{center}

\begin{center}
\begin{tabular}{|c||c|c|c|}
\hline
\multicolumn{1}{|c||}{\bf Collaboration } &
 \multicolumn{3}{c|}{
   $\nu \bar \nu\gamma\gamma$      }              \\   \hline
 & \multicolumn{1}{c|}{$\int {\cal L} d\tau$ , {\it pb}$^{-1}$} &
\multicolumn{1}{c|}{N, events}
&   \multicolumn{1}{c|}{$M_{\gamma\gamma}$, GeV}  \\   \hline
ALEPH  & 43 & 1 & 58.5$\pm$1.9                      \\   \hline
DELPHI & 40 & 0 & ---                               \\   \hline
L3     & 40 & 0 & ---                               \\   \hline
OPAL   & 43 & 0 & ---                               \\   \hline
\end{tabular}
\end{center}

\begin{center}
\begin{tabular}{|c||c|c|c|}
\hline
\multicolumn{1}{|c||}{\bf Collaboration } &
 \multicolumn{3}{c|}{
   $q \bar q \gamma\gamma$      }              \\   \hline
 & \multicolumn{1}{c|}{$\int {\cal L} d\tau$ , {\it pb}$^{-1}$} &
\multicolumn{1}{c|}{N, events}
&   \multicolumn{1}{c|}{$M_{\gamma\gamma}$, GeV}  \\   \hline
ALEPH  & 43 & 0 & ---                              \\  \hline
DELPHI & 40 & 0 & ---                               \\   \hline
L3     & 13 & 0 & ---                               \\   \hline
OPAL   & 43 & 1 & 59.5$\pm$0.2                      \\    \hline
\end{tabular}
\end{center}

	In the theoretical calculations using experimental data from
all four detectors --- ALEPH, DELPHI, L3 and OPAL, the value
of the physical efficiency is taken equal to unity.

	To determine the extreme values of the coupling constants
of the maximum likelihood method has been used \cite{9}.

\begin{eqnarray}
L(\bar x)\,=\,\prod_{i=1}^{5}\,P(N_i^{exp},\,N_i^{th}(\bar x))
\end{eqnarray}
where
\begin{eqnarray*}
P(n,\mu)\,=\,\frac{\mu^ne^{-\mu}}{n!}, \qquad\qquad n\,=\,0,1,\ldots.
\end{eqnarray*}
If the background processes with average $\mu_B$  taken into
account, this formula has some modifications:
\begin{eqnarray*}
P(n,\mu)\,=\,\frac{(\mu+\mu_B)^ne^{-(\mu+\mu_B)}}{n!}\,,\qquad\qquad n\,=\,0,
1,\ldots.
\end{eqnarray*}

	For the sake of application of the MINUIT \cite{13} we will use
below $\bar L\,=\,-\,{\it ln}L$ instead of value L.

\begin{center}
{\it Process $e^+e^-\,\to\,\mu^+\mu^-\gamma\gamma$.}
\end{center}

	The number of events of main process (1) is equal to:
\begin{itemize}
\item  in the case of scalar resonance:
\begin{eqnarray}
N_1^{th}\,&=&\,(0.713g_{\gamma\gamma}^2\,-\,4.25\cdot 10^{-3}
g_{\gamma\gamma}g_{\gamma z1}\,+\,    \nonumber  \\
&& 333.82g_{\gamma z1}^2\,+\,93.956g_{\gamma z2}^2\,-\,
6.429\cdot 10^{-4}g_{\gamma\gamma}g_{zz1}\,-\,  \nonumber  \\
&& 6.56\cdot 10^{-5}g_{\gamma\gamma} g_{zz2}\,-\,
0.478 g_{\gamma z2}g_{zz3}\,-\,0.235 g_{\gamma z1}
g_{zz2}\,-\,   \nonumber   \\
&& 4.605 g_{\gamma z1}g_{zz1}\,+\,3.552g_{zz1}^2\,+\,
0.443 g_{zz1}g_{zz2}\,+\,    \nonumber  \\
&& 0.0405g_{zz2}^2\,+\,0.134g_{zz3}^2)\,Br\,(\,R\,\to\,\gamma\gamma)
\end{eqnarray}
\item in the case of pseudoscalar resonance:
\begin{eqnarray}
N_1^{th}\,&=&\,(0.210g_{\gamma\gamma}^2\,-\,1.015\cdot 10^{-3}
g_{\gamma\gamma}g_{\gamma z2}\,+\,    \nonumber  \\
&& 333.82g_{\gamma z1}^2\,+\,93.956g_{\gamma z2}\,-\,
6.93\cdot 10^{-5}g_{\gamma\gamma}g_{zz3}\,-\,  \nonumber  \\
&& 0.478 g_{\gamma z2}g_{zz3}\,-\,0.235 g_{\gamma z1}
g_{zz2}\,-\,   \nonumber   \\
&& 4.605 g_{\gamma z1}g_{zz1}\,+\,3.552g_{zz1}^2\,+\,
0.443 g_{zz1}g_{zz2}\,+\,    \nonumber  \\
&& 0.0405g_{zz2}^2\,+\,0.134g_{zz3}^2)\,Br\,(\,R\,\to\,\gamma\gamma)
\end{eqnarray}
\end{itemize}
where the total integrated luminosity is $\int {\cal L} d\tau\,=\,166$
{\it pb}$^{-1}$ and $\sqrt{s}\,=\,91.2$ GeV. The average
value for the background is $ N_B\,=\,0.2$, and the experimental
number of events is $N_{exp\,1}\,=\,5$.

\begin{center}
{\it Process  $e^{+}e^{-}\,\to\,\nu\bar\nu\gamma\gamma$.}
\end{center}

	The number of events of this process is equal to:
\begin{eqnarray}
N_2^{th}\,&=&\,(9.904\cdot 10^{-3}g_{\gamma z1}^2\,+\,2.497\cdot 10^{-3}
g_{\gamma z2}^2\,+\,1.866\cdot 10^{-8}g_{\gamma z2}g_{zz3}  \nonumber  \\
&& \,-\,
1.96\cdot 10^{-5}g_{\gamma z1}g_{zz2}\,-\,4.87\cdot 10^{-5}g_{\gamma z1}
g_{zz1}\,+\,30.046g_{zz1}^2\,       \nonumber   \\
&& +\,6.99g_{zz1}g_{zz2} \,+\,1.01g_{zz2}^2\,+\,
1.169g_{zz3}^2)\,Br(R\,\to\,\gamma\gamma)
\end{eqnarray}
where we used total integrated luminosity $\int {\cal L} dt\,=\,166$
{\it pb}$^{-1}$ and $\sqrt{s}\,=\,91.2$ GeV. The average value
for the background is $N_B\,=\,0$, and the experimental number
of events is $N_2^{exp}\,=\,1$.

	One should note that the cross section of this process for scalar
resonance is equal to that of the pseudoscalar one.

\begin{center}
{\it Process $e^+e^-\,\to\,q \bar q\gamma \gamma$.}
\end{center}

	The number of events of the this process is equal to:
\begin{itemize}
\item  in case of scalar resonance:
\begin{eqnarray}
N_3^{th}\,&=&\,(13.38 g_{\gamma\gamma}^2\,-\,0.329 g_{\gamma\gamma}g_{\gamma
z1}
\,+\,
6267.232 g_{\gamma z1}^2\,+\,4666.786 g_{\gamma z2}^2  \nonumber  \\
&& \,-\,0.0404 g_{\gamma\gamma}g_{zz1}\,-\,
5.22\cdot 10^{-3}g_{\gamma\gamma}g_{zz2}\,-\,58.797g_{\gamma z2}g_{zz3}
\,-\, \nonumber  \\
&& 29.468 g_{\gamma z1} g_{zz2} \,-\,391.0 g_{\gamma z1}g_{zz1}\,+\,
95.493 g_{zz1}^2\,+\, \nonumber   \\
&& 20.016 g_{zz1}g_{zz2}
 \,+\,2.78 g_{zz2}^2 \,+\,3.684 g_{zz3}^2) Br(\,R\,\to\,\gamma\gamma)
\end{eqnarray}
\item  in case of pseudoscalar resonance:
\begin{eqnarray}
N_3^{th}\,&=&\,(9.964 g_{\gamma\gamma}^2\,-\,0.126 g_{\gamma\gamma}g_{\gamma
z2}
\,+\,
6267.232 g_{\gamma z1}^2\,+\,4666.786 g_{\gamma z2}^2\,  \nonumber \\
&& -\,0.054 g_{\gamma\gamma}g_{zz3}\,-\,
58.797g_{\gamma z2}g_{zz3}\,-\, \nonumber \\
&& 29.468 g_{\gamma z1}g_{zz2}   \,-\,391.0 g_{\gamma z1}g_{zz1}\,+\,
95.493 g_{zz1}^2\,+\, \nonumber   \\
&& 20.016 g_{zz1}g_{zz2}
 \,+\,2.78 g_{zz2}^2  \,+\,3.684 g_{zz3}^2) Br(\,R\,\to\,\gamma\gamma)
\end{eqnarray}
\end{itemize}
where we used the total integrated luminosity $\int {\cal L} d\tau\,=\,139$
{\it pb}$^{-1}$ and $\sqrt{s}\,=\,91.2$ GeV. The average value
for the background is $N_B\,=\,0$, and the experimental number
of events is $N_3^{exp}\,=\,1$.

\begin{center}
{\it $R$--resonance production in two photon annihilation at LEP energies}
\end{center}

	For the total integrated luminosity
$\int {\cal L} dt\,=\,43 $ {\it pb}$^{-1}$ and
$\sqrt{s}\,=\,91.2$ GeV one can obtain the following expression for the
number of events:
\begin{eqnarray}
N_4^{th}\,=\,1.56\cdot 10^3 g_{\gamma\gamma}^2\,
Br(R\,\to\,\gamma\gamma)
\end{eqnarray}

	The average value for the background is $N_B\,=\,0.32$,
and the experimental number of events is $N_4^{exp}\,=\,1$ \cite{2}.

\begin{center}
 {\it Process $e^+e^-\,\to\,Z\,\to\,3\gamma$. }
\end{center}

	The number of events of the process
\begin{eqnarray}
e^+e^-\,\to\,Z\,\to\,\gamma R(\,\to\,\gamma\gamma)
\end{eqnarray}
can be obtained from the following expression:
\begin{eqnarray}
N(e^+e^-\to Z\to 3\gamma)\,&=&\,L_{tot}\sigma(e^+e^-\to Z)
Br(Z\to \gamma R)Br(R\to \gamma\gamma)        \nonumber \\
\end{eqnarray}
where $\sigma(e^+e^-\to Z)\,=\,41.9$ nb \cite{9},
$L_{tot}\,=\,\int {\cal L} dt\,\simeq\,31$
 {\it pb}$^{-1}$. One can obtain the following number
of events process (18):
\begin{eqnarray}
N_5^{th}\,=\,6.62\cdot 10^4 (g_{\gamma z1}^2\,+\,g_{\gamma z2}^2)Br(R\,\to\,
\gamma\gamma)
\end{eqnarray}

	The average value for the background is $N_B\,=\,3.6$, and
the experimental number of events is $N_5^{th}\,=\,4$ \cite{7}.

	Substituting all values of $N_i^{th}$ and $N_i^{exp}$
in (11) and applying minimization routine MINUIT \cite{13}
, we obtain the extremum values of the coupling constants:
\begin{eqnarray}
&& g_{\gamma\gamma}\,=\,2.1\cdot 10 ^{-2} \qquad
g_{\gamma z1}\,=\,3.0\cdot 10^{-3} \qquad
g_{\gamma z2}\,=\,0.   \nonumber \\
&& g_{zz1}\,=\,0.220 \qquad
g_{zz2}\,=\,-\,0.751 \qquad
g_{zz3}\,=\,0.
\end{eqnarray}
The maximum number of events of main process (1) is equal to:
\begin{eqnarray}
N^{\pm}\,=\,0.13,
\end{eqnarray}
that gives an order of magnitude less events than that measured
experimentally.

 Note, that the dependence of the cross section for reaction (1), and,
correspondingly,
 of the number of the events from (12) and (13) on the minimal invariant mass
of final leptons is
 only logarithmic (see \cite{11}). Therefore choosing a different value for
 $(M_{\mu^+ \mu^-})_{min}$, e.g., 2 GeV, results in an increase of the cross
section and,
 consequently, of the number of the events not more than twice.  As is seen
from Table 1
 this is still considerably smaller than the experimental number of events.

It is worth mentioning the following fact. The estimates for the coupling
constant
(see (12)  --  (17) and (20))
contain the product of the constants squared by the branching of the
R resonance decay into two photons.

  The coupling constants for  the number of the events of process (1) in
final expressions (12) and (13)
are also multiplied by $Br(R\, \rightarrow \,\gamma \gamma)$, which
means that the analysis performed does not depend on the value of
 $Br(R \,\rightarrow \,\gamma \gamma)$.

	The main results of this Section are as follows. We have analyzed
the different experiments
a new R--resonance with a mass about 60 GeV, decaying
into $\gamma\gamma$ might be produced. Under general assumptions on the
interaction
of such resonance with photons and Z--bosons we obtained the upper
limits for the coupling constants of such type of interactions.
As a result we have 0.13 events of main reaction
(1) at the LEP energies. This number of events at least one
order of magnitude smaller than the experimentally observed number of
events with $m_{\gamma\gamma}\,=\,60$ GeV. So, one can may conclude
that these events are likely to be QED background and the explanation of
these events due to the existence of a new R resonance is almost
excluded.

	Nevertheless one can use the obtained the coupling constants
for upper limits in order to estimate the cross section of such
R resonance production at higher energies.

\section{\bf New R--resonance production at future $e^+e^-$ colliders}


	In this section
we consider the production this hypothetic resonance
in various processes at the energies above the $Z$ pole basing on
the upper limits of the coupling constants. We will study this
new R--resonance production in $e^+e^-$ and  $e\gamma$ collisions.
 On our calculations we assume, that the total
integrated luminosity will be $\int {\cal L} d\tau\,=\,10^3 {\it pb}^{-1}$.

\bc
{\it Processes $e^+e^-\,\to\,\mu^+\mu^-(\nu\bar\nu,q\bar q)\,\gamma\gamma$.}
\ec

    These processes are essentially the processes (1)  --  (3), but with a
different
$\sqrt{s}$. For the analitycal expressions for the cross--sections of these
processes see
\cite{11}. The background processes in these cases are:
\begin{eqnarray}
e^+ e^-\,\to\,\gamma^{\ast},Z^{\ast}\,\to\,\mu^+\mu^-(\nu\bar\nu,
q\bar q)\,\gamma\gamma
\end{eqnarray}
where both photons are produced due to hard radiation from the initial or
final charged fermions.
But this background could be separated
from the signal by demanding a large invariant mass of two photons
and taking into account different angular distributions of the
photons from the signal and background.

    The behaviour of the number of events versus $\sqrt{s}$
is presented in Figs. 1, 2 and 3. The corresponding number of events
for three final states is the following:
\begin{eqnarray}
N^{\pm}_{\mu\mu}\,=\,0.04\quad
N^{\pm}_{\nu\bar\nu}\,=\,0.33 \qquad \quad
N^{\pm}_{q\bar q}\,=\,1.3 \,\,\mbox{ for } \sqrt{s}\,=\,100\,\mbox{ GeV }
\nonumber \\
N^{\pm}_{\mu\mu}\,=\,15.2\quad
N^{\pm}_{\nu\bar\nu}\,=\,82.0 \quad
N^{\pm}_{q\bar q}\,=\,324.3 \,\,\mbox{ for } \sqrt{s}\,=\,200\,\mbox{ GeV }
\end{eqnarray}

\begin{center}
{\it Process $\, e\gamma \,\to\, eR\,(\,\to\gamma\gamma)$}
\end{center}

    The R--resonance can be produced in the following reaction too:
\begin{eqnarray}
e\gamma\,\to\,eR\,(\to\,\gamma\gamma)
\end{eqnarray}

	This process is described by two Feynman diagrams, that correspond
to the photon and $Z$--boson exchange at the $t$--channel. The analytical
expressions for the cross-section of this process Appendix can be found.

	 The behaviour of the number of events of process (25)
versus $\sqrt{s}$ is presented on fig. 4.

	 For process (25) the background is $e\gamma\,\to\,e\gamma$
with photon emission from the initial or final leptons. The signature of
process (25)
is the fact, that two photon invariance mass is equal to $60$ GeV, and
after some cut for this two photons invariance mass, the background process is
easily suppressed.

As one can see  from
fig.4 for $\sqrt{s}\,=\,100,\,200$ GeV the following
number of events can be obtained:
\begin{eqnarray}
N^{\pm}\,=\,737.0    \qquad\qquad \mbox{  for  } \sqrt{s}\,=\,100\, \mbox{GeV}
\\
N^{\pm}\,=\,1743.0    \qquad\qquad \mbox{  for  } \sqrt{s}\,=\,200\, \mbox{GeV}
\end{eqnarray}

%
%
%
%
%

\section{\bf CONCLUSIONS}

  The $e^+\, e^- \,\rightarrow \,l^+\,l^- \,\gamma \,\gamma$  events
 detected in $L3$ at $CERN$ have been
 analyzed proceeding from the general form of the interaction of
 scalar (pseudoscalar)
 resonance with photon and $Z$--boson.

 The analysis of the reactions
\begin{enumerate}
\item $e^+\,e^-\,\to\, \mu \bar \mu \gamma \gamma$,
\item $e^+\,e^-\,\to\, \nu \bar \nu \gamma \gamma$,
\item $e^+\,e^-\,\to\, q \bar q \gamma \gamma$,
\item $e^+\,e^-\,\to\,e^+\,e^-\,\gamma \gamma$ (two photon annihilation at LEP
energies),
\item $Z\,\to\, \gamma \gamma \gamma$,
\end{enumerate}
allowed one to obtain the constraints for the values of the effective constants
of the resonance interaction with photon and $Z$--boson
 ($g_{\gamma\gamma}$, $g_{\gamma z1}$, $g_{\gamma z2}$, $g_{zz1}$,
$g_{zz2}$, $g_{zz3}$).

 The limitations obtained for the coupling constants
bring us to the total number of the events of the reaction
 $e^+\, e^-\, \rightarrow \,\mu^+ \,\mu^- \,\gamma \,\gamma$
which by an order of magnitude is lower than in the
 $LEP-L3$ experiment (the case of the pseudoscalar resonance
has approximately the same number
 as the scalar one). This conclusion confirms our previous
results from \cite{11}.

	Nevertheless, we make some predictions of such
R--resonance production at future $e^+e^-$ colliders,
based on the obtained values of the coupling constants.

    There exist some promissing processes at the LEP--200 and VLEPP machines
where this hypothetic resonance could be detected.
We have calculated a possible number of events of this
resonance production for   $\int {\cal L} d\tau\,=\,10^3
\,{\it pb}^{-1}$. As it follows from these calculations for the case
of $\sqrt{s}\,=\,100$ GeV the number of events is very small (not greater than
unity).
But for $\sqrt{s}\,=\,200$ GeV one could expect from 15 events (for
the $\mu\mu\gamma\gamma$ final state) up to $\simeq$300 (for the $
q\bar q\gamma\gamma$ final state) and up to $\simeq\,2000$ in the
$e\gamma$ collisions.
 Thus the search of such type of
processes at $\sqrt{s}\,=\,200$ GeV (at LEP--200 and VLEPP)
could be used to confirm the existence of such resonance.

\vskip0.5cm
{\bf Acknowledgements.}

\noindent We thank Yu.M.Antipov, V.I.Borodulin, E.A. Kushnirenko, A.A.
Likhoded, A.K.Li\-kho\-ded,
V.F.Obraz\-tsov, A.V.Shvorob and  A.M.Zaitsev
 for fruitful
discussions.

\newpage
\begin{flushright}
{\bf Appendix}
\end{flushright}

    Here we consider the cross sections for the process:
\begin{eqnarray}
e\gamma \,\to\,e\, R(\,\to\,\gamma\gamma) \nonumber
\end{eqnarray}

    This process is described by
two Feynman diagrams.
The matrix elements corresponding to this diagrams are:
\begin{enumerate}
\item [1.] for scalar R:
\begin{eqnarray}
M\,&=&\,\varepsilon^{\alpha} \frac{g_{\gamma\gamma}}{m_R}
f^{\alpha\beta}L^{\beta}\frac{1}{q^2}\,+\,\varepsilon^{\alpha}
\frac{g_{\gamma z1}}{m_R}f^{\alpha\beta}\frac{1}{Z}d^{\beta\sigma}
K^{\sigma}\,+\,              \nonumber \\
&& \varepsilon^{\alpha}\frac{g_{\gamma z2}}{m_R}t^{\alpha\beta}
\frac{1}{Z}d^{\beta\sigma} K^{\sigma} \nonumber
\end{eqnarray}
\item [2.] for pseudoscalar R:
\begin{eqnarray}
M\,&=&\,\varepsilon^{\alpha} \frac{g_{\gamma\gamma}}{m_R}
t^{\alpha\beta}L^{\beta}\frac{1}{q^2}\,+\,\varepsilon^{\alpha}
\frac{g_{\gamma z1}}{m_R}f^{\alpha\beta}\frac{1}{Z}d^{\beta\sigma}
K^{\sigma}\,+\,              \nonumber \\
&& \varepsilon^{\alpha}\frac{g_{\gamma z2}}{m_R}t^{\alpha\beta}
\frac{1}{Z}d^{\beta\sigma} K^{\sigma} \nonumber
\end{eqnarray}
\end{enumerate}
where the following notations are introduced:
\begin{eqnarray*}
&& Z = (q^2-m_Z^2)+im_Z\Gamma_Z, \quad  \\
&& f^{\alpha\beta} = (qk)g^{\alpha\beta}-k^{\beta}q^{\alpha}, \,
 d^{\alpha\beta} = g^{\alpha\beta}-\frac{q^{\alpha}
q^{\beta}}{m_Z^2},
t^{\alpha\beta}\,=\,\varepsilon^{\alpha\beta\lambda\sigma}q^{\lambda}
k^{\sigma},   \\
&& L^{\alpha} = e \bar u(k_1)\gamma^{\alpha}u(-k_2),
 K^{\alpha} = \bar u(k_1)\gamma^{\alpha}(c_v+c_a\gamma^5)
u(-k_2).
\end{eqnarray*}
where
 $e$ is the electron charge (in electron charge units);
 $c_v,c_a$ are the vector and axial coupling constants of initial
(final) fermions with $Z$--boson.

    One can obtain the following results
(for the massless fermions):
\begin{itemize}
\item
[---] in the case of scalar resonance:
\begin{eqnarray}
\frac{d\sigma}{d(cos\theta)}\,&=&\,\frac{1}{64\pi m_R^2}
(1-\delta)^2 \frac{1+2cos\theta +cos^2\theta +
\frac{4}{(1-\delta)^2}}{(1-cos\theta +\frac{2}
{a_1-a_2})^2+\frac{4g}{(a_1-a_2)^2}} \cdot  \nonumber  \\
&& (\,-\,c_v e g_{\gamma\gamma}g_{\gamma z1}\cdot (2-2cos\theta +\frac{4}
{a_1-a_2})\,+\,       \nonumber \\
&& (g_{\gamma z1}^2+g_{\gamma z2}^2)(c_v^2+c_a^2)(1-cos\theta))
\,+\, \nonumber \\
&& \frac{e^2g_{\gamma\gamma}^2}{16\pi m_R^2}
\frac{1+\frac{1}{4}(1-\delta)^2(1+cos\theta)^2}
{1-cos\theta+\frac{c}{1-\delta}(1+\frac{1+\delta}{1-\delta}
cos\theta)}     \nonumber
\end{eqnarray}
\item [---] in the case of pseudoscalar resonance one:
\begin{eqnarray}
\frac{d\sigma}{d(cos\theta)}\,&=&\,\frac{1}{64\pi m_R^2}
(1-\delta)^2 \frac{1+2cos\theta +cos^2\theta +
\frac{4}{(1-\delta)^2}}{(1-cos\theta +\frac{2}
{a_1-a_2})^2+\frac{4g}{(a_1-a_2)^2}} \cdot  \nonumber  \\
&& (\,-\,c_v e g_{\gamma\gamma}g_{\gamma z2}\cdot (2-2cos\theta +\frac{4}
{a_1-a_2})\,+\,       \nonumber \\
&& (g_{\gamma z1}^2+g_{\gamma z2}^2)(c_v^2+c_a^2)(1-cos\theta))
\,+\, \nonumber \\
&& \frac{e^2g_{\gamma\gamma}^2}{16\pi m_R^2}
\frac{1+\frac{1}{4}(1-\delta)^2(1+cos\theta)^2}
{1-cos\theta+\frac{c}{1-\delta}(1+\frac{1+\delta}{1-\delta}
cos\theta)}     \nonumber
\end{eqnarray}
\end{itemize}
where
\begin{eqnarray}
&& \delta\,=\,\frac{m_R^2}{s}\,,\, \qquad a_1\,=\,\frac{s}{m_Z^2}\,,\,\qquad
a_2\,=\,\frac{m_R^2}{m_Z^2}    \nonumber \\
&& g \,=\,\frac{\Gamma^2}{m_Z^2}\,,\,\qquad c\,=\,\frac{m_e^2}{s} \nonumber
\end{eqnarray}

\vspace*{3cm}

\begin{flushright}
{\it Received February, 18, 1994.}
\end{flushright}
\newpage
\begin{figure}[t]
\epsffile{csmu.ps 10cm 10cm}
\caption{ Total number of events for process $e^+e^-\,\to\,\mu \mu
\gamma\gamma$
($\int {\cal L} d\tau\,=\,10^3$ {\it pb}$^{-1}$).}
\end{figure}
%
\begin{figure}[b]
\epsffile{csnu.ps 10cm 10cm}
\caption{ Total number of events for process
$e^+e^-\,\to\,\nu\bar\nu\gamma\gamma$
($\int {\cal L} d\tau\,=\,10^3$ {\it pb}$^{-1}$).}
\end{figure}
\vspace{1cm}
\newpage
\begin{figure}[h]
\epsffile{csq.ps 10cm 10cm}
\caption{ Total number of events for process $e^+e^-\,\to\,q\bar q\gamma\gamma$
($\int {\cal L} d\tau\,=\,10^3$ {\it pb}$^{-1}$).}
\end{figure}
\vspace{1cm}
\begin{figure}[b]
\epsffile{cseg.ps 10cm 10cm}
\caption{ Total number of events for process $e \gamma\,\to\,e R\,(\,\to
\gamma\gamma) $
($\int {\cal L} d\tau\,=\,10^3$ {\it pb}$^{-1}$).}
\end{figure}
\vspace{1cm}
%

\end{document}